\documentclass[12pt,preprint]{aastex}


\usepackage{epstopdf, graphicx, natbib}
\bibpunct{(}{)}{,}{a}{}{;}
\usepackage{times}
\usepackage{rotating,lscape}
\usepackage{epsfig}
\usepackage{graphics,graphicx}
\usepackage{epstopdf}

\def\chandra{{\it Chandra}~}
\def\xmm{{\it XMM-Newton}~}

\def\hst{{\it Hubble Space Telescope}~}
\def\HST{{\it HST}~}
\def\swift{{\it Swift}~}

\newcommand{\ignore}[1]{}

\def\lya{\ifmmode {\rm Ly}\alpha~ \else Ly$\alpha$~\fi}
\def\lyan{\ifmmode {\rm Ly}\alpha \else Ly$\alpha$\fi}
\def\lyb{\ifmmode {\rm Ly}\beta~ \else Ly$\beta$~\fi}
\def\lyg{\ifmmode {\rm Ly}\gamma~ \else Ly$\gamma$~\fi}
\def\civ{\ifmmode {\rm C}\,{\sc iv}~ \else C\,{\sc iv}~\fi}
\def\civn{\ifmmode {\rm C}\,{\sc iv}~ \else C\,{\sc iv}\fi}
\def\cvi{\ifmmode {\rm C}\,{\sc vi}~ \else C\,{\sc vi}~\fi}
\def\cvin{\ifmmode {\rm C}\,{\sc vi} \else C\,{\sc vi}\fi}

\def\ciii{C\,{\sc III}~}

\def\cv{C\,{\sc v}~}

\def\heii{He\,{\sc ii}~}

\def\siiii{Si\,{\sc III}~}

\def\siiv{Si\,{\sc IV}~}

\def\oiii{{{\rm O}\,{\sc iii}~}}

\def\oiv{{{\rm O}\,{\sc iv}~}}

\title{Space Telescope and Optical Reverberation Mapping Project. VII.\
   Understanding the Ultraviolet Anomaly in NGC\,5548 with X-Ray Spectroscopy}

 \author{
 S.~Mathur\altaffilmark{1,2}, 
 A.~Gupta\altaffilmark{2,3}, 
 K.~Page\altaffilmark{4}, 
 R.W.~Pogge\altaffilmark{1,2}, 
 Y.~Krongold\altaffilmark{5}, 
 M.R.~Goad\altaffilmark{4}, 
 S.M.~Adams\altaffilmark{1,6}, 
 M.D.~Anderson\altaffilmark{7}, 
  P.~Ar\'{e}valo\altaffilmark{8}, 
  A.J.~Barth\altaffilmark{9}
% ...
} 

\altaffiltext{1}{\ignore{OSU}Department of Astronomy, The Ohio State
  University, 140 W 18th Ave, Columbus, OH 43210, USA}
\altaffiltext{2}{\ignore{CCAPP}Center for Cosmology and AstroParticle
  Physics, The Ohio State University, 191 West Woodruff Ave, Columbus,
  OH 43210, USA} \altaffiltext{3}{\ignore{ColsState}Department of
  Biological and Physical Sciences, Columbus State Community College,
  Columbus, OH 43215, USA} \altaffiltext{4}{\ignore{Leicester}Department
  of Physics and Astronomy, University of Leicester, Leicester, LE1 7RH,
  UK} \altaffiltext{5}{\ignore{UNAM}Instituto de Astronomia, Universidad
  Nacional Autonoma de Mexico, Cuidad de Mexico, Mexico}
\altaffiltext{6}{\ignore{Caltech}Cahill Center for Astrophysics,
  California Institute of Technology, Pasadena, CA 91125, USA}
\altaffiltext{7}{\ignore{GSU}Department of Physics and Astronomy,
  Georgia State University, 25 Park Place, Suite 605, Atlanta, GA 30303,
  USA} \altaffiltext{8}{\ignore{Valapaiso}Instituto de F\'{\i}sica y
  Astronom\'{\i}a, Facultad de Ciencias, Universidad de Valpara\'{\i}so,
  Gran Bretana N 1111, Playa Ancha, Valpara\'{\i}so, Chile}
\altaffiltext{9}{\ignore{UCI}Department of Physics and Astronomy, 4129
  Frederick Reines Hall, University of California, Irvine, CA 92697,
  USA}
% ...

\begin{document}
   
\begin{abstract}
  During the Space Telescope and Optical Reverberation Mapping Project
  (STORM) observations of NGC 5548, the continuum and emission-line
  variability became decorrelated during the second half of the 6-month-long 
  observing campaign. Here we present \swift and \chandra X-ray
  spectra of NGC 5548 obtained as a part of the campaign.  The \swift
  spectra show that excess flux (relative to a power-law continuum) in
  the soft X-ray band appears before the start of the anomalous
  emission-line behavior, peaks during the period of the anomaly, and
  then declines. This is a model-independent result suggesting that the
  soft excess is related to the anomaly. We divide the \swift data into
  on- and off-anomaly spectra to characterize the soft excess via
  spectral fitting.  The cause of the spectral differences is likely
  due to a change in the intrinsic spectrum rather than to
  variable obscuration or partial covering.  The \chandra spectra have
  lower signal-to-noise ratios, but are consistent with the \swift data. Our
  preferred model of the soft excess is emission from an optically
  thick, warm Comptonizing corona, the effective optical depth of which
  increases during the anomaly. This model simultaneously explains all
  three observations: the UV emission-line flux decrease, the
  soft-excess increase, and the emission-line anomaly.

\end{abstract}

\section{Introduction}

The Space Telescope and Optical Reverberation Mapping (STORM) project
intensively monitored the well-known active galactic nucleus (AGN)
NGC~5548. As a part of this project, NGC~5548 was observed with the \hst
({\it HST}) in 2014 for 180 days with daily cadence, obtaining 171
usable epochs. The source was also monitored with \swift and in the
optical with ground-based observations. In addition, we observed the
source with \chandra four times during the {\it HST} observing
campaign. The goal of the STORM project was to perform velocity-resolved
reverberation mapping (RM) of the optical and ultraviolet (UV) emission
lines with fine time sampling, long duration, and high signal-to-noise
ratio (S/N) spectra. The multiwavelength continuum observations were
performed to probe the structure of the accretion disk and to track
changes in the ionizing continuum of the source.

The \HST (UV), \swift (X-ray), and ground-based (optical) continuum and
spectroscopic observations are presented in Papers I--V (De Rosa et
al. 2015; Edelson et al. 2015; Fausnaugh et al. 2016; Goad et al. 2016;
and Pei et al. 2017, respectively). Reverberating-disk models for NGC
5548 are presented in Paper VI (Starkey et al. 2017). As noted in Paper
I and discussed in detail in Paper IV (Goad et al. 2016), an anomalous
behavior of the broad UV emission lines was observed during the
campaign. For most of the campaign, the broad UV emission lines
responded to changes in the UV continuum, as generally expected for
broad-line reverberation. However, there was a period of 60--70 days
during the latter half of the campaign when the UV lines did not
reverberate with the UV continuum. This was also accompanied by a
significant drop in the fluxes and equivalent widths of UV and optical
emission lines, to varying degrees.

Such an ``anomaly,'' when the continuum and emission-line variability
became decoupled, was not previously observed in RM campaigns and
demands explanation.  Moreover, understanding the origin of the anomaly
is critical to the robustness of the RM technique, since
it depends on the observed continuum flux being a good proxy for the
unobserved extreme-UV (EUV) ionizing continuum.  Here we present X-ray spectra
obtained as part of AGN STORM that provide important clues to the
origin of the anomaly. In \S 2 we present analysis of \swift
spectra. The \chandra observations and spectral analysis are presented
in \S 3.  In \S 4 we discuss how the appearance of a soft-X-ray excess a
few days before the anomaly may clarify its origin. Although NGC 5548
was observed intensively with \xmm in the years prior to the AGN STORM
campaign (Kaastra et al. 2014; Mehdipour et al. 2015, 2016; Cappi et
al. 2016), a detailed comparison with those data is beyond the scope of
this paper. Here we focus on \swift and \chandra data obtained during
our campaign.

\section {{\it Swift} Observations} 

\subsection {Spectra and Analysis}

Descriptions of the \swift observations, data reduction, and time-series
analysis are presented in Paper II. Here we present a time-resolved
spectral analysis. In Figure 1, we show the $0.3$--$10.0$ keV spectra in
9 time bins, from pre-anomaly (days $1$--$54$) to post-anomaly (days
$150$--$170$). The exposure times for each period are given in
Table~1. We see that the hard X-ray continuum ($2$--$10.0$ keV) is
constant over the period of the observation, but the soft X-ray flux
(below $\sim 0.8$ keV) increases from the pre-anomaly period (days
$1$--$54$) to days $55$--$75$, peaks during days $75$--$85$, and then
fades during days $85$--$100$.  This is a model-independent result,
suggesting that the soft excess is likely related to the anomaly. {\bf
  In Figure 2, we have reproduced Figure 1e from Paper IV; the black
  points show the percentage difference in the \civ flux.  The red
  points show the percentage difference in count rate at $\approx 0.55$
  keV. This again shows that the soft excess increases on days
  $55$--$75$, just before the start of the anomaly (on day $\sim75$),
  peaks during days $75$--$85$, and then fades. This also shows that
  there is some delay of about $20$ days between the period of high
  soft-excess and the period of the anomaly}.

In order to perform the spectral modeling and to quantify the soft
excess, we divided the \swift observations into two parts, which we call
pre-anomaly and anomaly spectra (days $0.4$ to $54.5 =$ JD
$2,456,690.4189$ to $2,456,744.5088$, and days $55.4$ to $84.9 =$ JD
$2,456,745.3676$ to $2,456,774.9165$, respectively). A Galactic column
density of $N_{\rm H}=1.69 \times 10^{20}$ cm$^{-2}$ (Dickey \& Lockman 1990)
was included in all the models. We simultaneously fitted both the  spectra
with an absorbed power-law model, adding a blackbody (BB)
component to parameterize the soft excess. The parameters of the
power-law slope (photon index $\Gamma$) and normalization were tied
for the two spectra (as justified by Fig. 1), but the intrinsic
absorption and BB parameters were allowed to vary. The fit showed that
the BB temperature is the same in the two spectra, so we tied the
temperature and fitted the spectra again. The resulting fit was good
($\chi^2_{\nu}=1.16$ for 1001 degrees of freedom).

In their model of the entire \xmm observing campaign, Cappi et
al. (2016) found an additional scattered soft X-ray component dominated
by narrow emission lines, with 8\% of the total soft X-ray flux. This
soft component was constant over the whole campaign, so it cannot be
responsible for the variable soft excess we see here.  Nonetheless, we
added a similar component to our model ({XSPEC model \bf apec}) with
flux as in Cappi et al. (2016) and found no improvement to the fit
($\chi^2_{\nu}=1.16$ for 1001 degrees of freedom).  The best-fit spectra
are shown in Figure 3 and the model parameters are given in Table 2
along with the flux in the soft X-ray excess after correcting for
absorption. The soft excess is significantly stronger in the anomaly
spectrum. The intrinsic spectrum (without the instrument response) is
shown in the bottom panel of Figure 3; the absorbed power law, BB, and
scattered emission-line components are shown as dotted lines.

While a BB component describes the soft excess well, ``warm
Comptonization,'' in which seed photons from the accretion disk are
Compton upscattered by an optically thick hotter corona is likely a
more realistic model. Thus, we also tried to fit the soft excess with a
warm Comptonization model ({\bf compTT} in XSPEC), together with an
absorbed power law. We fixed the model parameters (seed photon
temperature, corona temperature, and optical depth) to the parameters in
Medhipour et al. (2015), allowing only the normalization to vary between
the pre-anomaly and anomaly spectra. The resulting fit was worse
($\chi^2_{\nu}=1.33$ for 1015 degrees of freedom; $\Delta \chi^2=189$
compared to the BB fit) and significant negative residuals were observed
below 1 keV; in particular the warm Comptonization model did not
adequately fit the spectral turnover below 0.5 keV. We could improve the
fit by adding another cold absorber at the source (best fit
$N_{\rm H}=(4.7\pm0.7) \times 10^{20}$ cm$^{-2}$), resulting in
$\chi^2_{\nu}=1.2$ for 1014 degrees of freedom. Hence, if warm
Comptonization is the correct description of the soft excess, it will
have to lie behind this absorbing medium. The best-fit normalizations
are given in Table 3 and the fit is shown in Figure 4; as expected, the
normalization is significantly higher during the anomaly. Alternatively,
the change in soft excess could be a result of a changing optical depth of
the corona, as shown by Page et al. (2004).  We therefore fixed the
normalization to the pre-anomaly value and allowed only the optical depth to
vary. As expected, we find significantly higher optical depth during the
anomaly (Table 3). A warm Comptonized corona with variable optical depth
is our preferred model for reasons discussed in \S4.2.

{\bf  When we fit the warm
  Comptonization model to the post-anomaly spectrum (days $150$--$170$),
  the normalization is $29\pm 2$ as compared to $22 \pm 1$ in the
  pre-anomaly spectrum and $50\pm 2$ during the anomaly. At late times,
  the soft excess has dropped significantly and is close to the excess
  in the early phase.  While there is some delay with respect to the
  emission line anomaly and the continuum transitions are not very sharp
  (as we see in Fig. 1), this strengthens our assertion that the
  increase in the soft-excess is likely the cause of the anomaly. For
  the rest of the paper we focus only on spectra before and during the
  anomaly.  }

\subsection {Alternative Models}

It is possible that the observed spectral shape (the ``soft excess'') is
an artifact of a partially covering cold absorber.  We therefore tried
to fit the soft excess with a partial covering power-law model. The
best-fit absorption column densities in the two spectra were found to be
the same within the uncertainties, so we refitted the two spectra after tying
the column densities. As expected, the absorber covered the continuum
source more during the pre-anomaly period (covering fraction
$0.91\pm0.01$) and less during the anomaly (covering fraction
$0.75\pm0.02$). However, this fit is worse ($\chi^2_{\nu}=1.4$ for 1003
degrees of freedom) than the BB model, with obvious residuals in the
soft X-ray spectrum. As seen in Figure 5, the partial covering absorber
model does not adequately account for the excess flux in the soft X-ray
band in the anomaly spectrum, or the spectral turnover below $\sim 2$
keV. Thus, the {\it dominant} parameter describing the change in soft
excess between pre-anomaly and anomaly spectra cannot be the covering
factor.

Alternatively, the apparent soft ``excess'' could result from the
recovery of the soft X-ray spectrum through a warm absorber. NGC~5548 is
known to have an X-ray warm absorber, detected even in low-resolution
spectra (e.g., Nandra et al. 1993; Mathur et al. 1995). We therefore
tried to fit the \swift spectra with an absorbed power-law model
modified by a warm absorber. We used PHASE (Krongold et al. 2003) to
model the warm absorber. The parameters of this model are ionization
parameter $U$, total column density $N_{\rm H}$, velocity $v$, and the
microturbulent velocity $\sigma$. Given the low resolution of the
spectrum, we fixed $v$ to match the galaxy redshift and $\sigma$ was
fixed to 200~km~s$^{-1}$. Thus, the free parameters of the warm absorber
model were $U$ and $N_{\rm H}$.  Once again, we fitted the two spectra
simultaneously, keeping the power-law parameters tied. The fit was good
($\chi^2_{\nu}=1.1$ for 1003 degrees of freedom) and the results of this
fit are given in Table 4. The absorber $N_{\rm H}$ was found to be
similar in the two spectra, and as expected, the warm absorber in the
anomaly spectrum is more ionized (higher $U$), leading to the apparent
excess in the soft X-ray band.  The warm absorber column density,
however, is unusually large ($\log N_{\rm H} ({\rm
  cm^{-2}}) \approx 22.2$). The warm absorber column density in NGC~5548 has
varied from about $\log N_{\rm H} ({\rm cm^{-2}})=20.3$ to $\log N_{\rm
  H} ({\rm cm^{-2}})=21.7$ (Mehdipour et al. 2015). While a significant
increase in the column density is possible, it is also possible that the
column density is actually lower, and there is an additional soft-excess
component; we cannot distinguish between these two models.  Warm
absorbers with low ionization parameter are clearly related to UV
absorption lines (e.g., Mathur et al. 1994,,1995,,1998; Monier et al. 2001;
Krongold et al. 2003, 2005, 2007; Kaspi et al. 2004), so variability of
the UV absorber (Kriss et al., in prep.) may help distinguish between the
two possibilities. As noted below in \S 4.2, the warm absorber model
alone cannot explain the anomaly, so a variable warm absorber is
unlikely to be the correct model of the soft excess.

As noted in \S 1, NGC~5548 was monitored intensively with \xmm in
2013--2014. The source appeared in a highly obscured state then (Kaastra
et al. 2014), with a column density $\sim 10^{23}$ cm$^{-2}$. In
our STORM campaign, the absorption column density was over an order of
magnitude lower ($\sim 10^{22}$ cm$^{-2}$). A detailed spectral
analysis of \xmm observations of NGC~5548 is presented by Cappi et
al. (2015), who modeled the spectra with six different components: a
power-law continuum; a cold reflection component; a soft thermal
Comptonization emission model; a scattered emission-line component; a
warm absorber; and up to two high column density partial covering
``obscurers.'' Given the S/N of our \swift observations, it is not
possible to fit the spectra with such a complex model and deduce
meaningful information. Moreover, our interest is to understand the {\it
  difference} between the pre-anomaly and anomaly spectra.  It is
possible that the observed changes in the soft excess are caused by a
combination of changes in multiple components. We cannot constrain these
multiple components; instead, we have looked for a dominant model that
describes the soft-excess variability. During the \xmm campaign, the
dominant variability component in the $0.3$--$0.8$ keV range of soft
excess was the covering fraction of the obscurer as reported by
Mehdipour et al. (2016), but Cappi et al. (2015) show that the
normalization of the Comptonization component was also important. 

In our STORM campaign, the soft excess can be modeled as a BB, a warm
absorber, a warm Comptonizing medium behind a thin veil of matter, or a
combination of all these models; the \swift spectra cannot discriminate
among these models. However, the dominant variability is {\it not}
caused by the covering fraction {or the column density} of an
absorber. As we discuss further in \S 4, the warm Comptonization model
naturally explains several aspects of the UV anomaly, so this is our
preferred model. While a change in the normalization of the warm
Comptonization model can adequately describe the spectral difference
between the pre-anomaly and anomaly spectra, our preferred model is of
the change in the effective optical depth (\S 4). Thus, we see that the
{\it shape} of the X-ray continuum changed during the anomaly phase, not
just the normalization; this was clear from the model independent
spectra shown in Figure 1, and the spectral modeling confirms the
same. Recently, Gardner \& Done (2017) studied optical/UV variability of
NGC~5548 and argued that a soft excess is required to understand the
observed continuum lags; the observation of a soft excess with {\swift}
and \chandra (\S 3) is consistent with this expectation.

\section{\chandra Observations and Analysis}

As a part of the \HST campaign, we also observed NGC~5548 with the
\chandra Low Energy Transmission Grating (LETG) and ACIS-S on four
occasions from Feb. 2014 to June 2014 for 5~ks each. The observation
details are given in Table 5. The LETG was placed in front of the
detector to avoid pile-up; obtaining high-resolution grating spectra was
not the goal of these short 5~ks exposures.  We analyzed the data using
standard CIAO tools (version 4.7 and $\it caldb$ version 4.6.7). All the 
observations were reprocessed using the \textit{Chandra repro} task,
which results in enhanced data quality and better calibration. We
extracted the zeroth-order source and background spectra using the CIAO
tool \textit{specextract}, which also builds proper response (RMFs) and
effective area (ARF) files required for analysis.  We analyzed the
spectra using both \textit{XSPEC} and the CIAO fitting package
\textit{Sherpa}.  We binned the spectra to 25 counts minimum per channel
using \textit{ftool grppha}.

It was clear that in the hard X-ray band (2--8 keV) the spectra are very
similar, but at softer energies they show differences, similar to what
we found in \swift spectra. Thus, we fit the \chandra spectra with an
absorbed power-law model, as we did for \swift spectra, with the results
shown in Figure 6. Once again we see that there is a clear soft excess
in \chandra observations II (day $57 =$ JD $2,456,747$) and III (day $93
=$ JD $2,456,784$), which were taken just before and during the
anomaly. Observations I and IV took place when the source was in its
normal state, and they show no soft excess. The appearance of the soft
excess in observations II and III once again suggests that it may be
related to the anomaly.

We fit the \chandra spectra with the same series of models discussed
above, primarily to determine if they are consistent with the \swift
results despite their lower S/N.  The models which fit the \swift data
well also fit the the \chandra data well, and the partial covering
power-law model is a poor model of the \chandra spectra as well
($\chi^2_{\nu}=2.32$ for $\nu =63$ degrees of freedom), with the fit
yielding a covering fraction of unity.

\section{Results}
\subsection{Soft-Excess}
Ever since the discovery of soft X-ray excesses (Singh, Garmire, \& Nousek 
1985; Arnaud et al. 1995), there has been a debate about their origin and
physical nature. The possible explanations have narrowed down to 
(1) reflection of the hard X-ray source by the accretion disk
(e.g., Crummy et al. 2006); (2) an additional Comptonizing medium around
the accretion disk (e.g., Ross, Fabian, \& Mineshige 1992); 
or (3) thermal emission from
an accretion disk.  Understanding the nature of the soft excess is
important because of its potentially large luminosity and because it is
an integral part of the accretion process. Though no correlation has
been found between the strength of the soft-excess and the black hole
mass or its luminosity (e.g., Bianchi et al. 2009), multiwavelength
studies have revealed a possible correlation of the UV slope with the
soft-excess strength and shape (e.g., Walter \& Fink 1993; Atlee \&
Mathur 2009). From the multiwavelength campaign studying Mrk~509,
Mehdipour et al. (2011) found that the soft X-ray excess is correlated
with the thermal optical-UV emission from the accretion disk and is not
correlated with the $2$--$10$ keV X-ray power law. This favors
Comptonization of UV/optical photons by a hot plasma for the origin of
the soft excess.


In NGC~5548, the soft excess was previously detected in 2000 during an
unobscured period, but the source was heavily absorbed during 2013
(Kaastra et al. 2014) and the soft-excess was modeled as a Comptonized
corona (Mehdipour et al. 2015). The soft excess observed during our
campaign is well modeled as a BB (an optically thick corona
would emit like a BB), but the anomaly is better explained by
the warm Comptonization model, as discussed below, so this is our
preferred explanation.

\subsection{Understanding the UV Anomaly}

The X-ray spectra provide a possible explanation for the UV
anomaly. First, the X-ray spectra rule out variable absorption as the
cause of the anomaly. If anything, the absorption was lower during the
anomaly.  It is possible that the EUV ionizing continuum source was
obscured, while the X-ray source was not, but that is unlikely since the
X-ray continuum source size in AGNs appears to be smaller than that of
the UV/EUV (e.g.,  Mosquera et al. 2013). Partial covering of the
continuum fits the data poorly and is unlikely to be the cause of
the anomaly.

The second important fact is that the UV emission-line flux decreased
during the anomaly. This suggests that the EUV ionizing continuum flux
decreased during the anomaly (as discussed in Paper IV; the ionization
potential of \ciii is 47.9 eV and that of \siiii is 33.5 eV). This
cannot be explained by a change in the warm absorber because the warm
absorber in the anomaly spectrum is more ionized (higher $U$), requiring
an increase in the EUV continuum during the anomaly. Thus, our preferred
scenario is that of the intrinsic change in the soft X-ray spectrum.

The observed changes can be naturally explained by the warm
Comptonization model. In this model, the UV disk photons are Compton
upscattered to soft X-ray energies by the optically thick corona. In a
normal situation, where the UV and soft X-ray fluxes are correlated,
higher UV flux leads to more input photons for Comptonization, so more
soft X-rays; we can describe this as a change in the normalization of
the model. What we have, however, is the opposite situation during the
anomaly: the EUV flux decreases, while the soft X-ray flux
increases. Perhaps the EUV photons are depleted from the flux {\bf seen
  by the BLR}, but are Comptonized into soft X-rays. This could be due
to either a true change in the optical depth or an increase in the
region covering the UV/EUV-emitting region, which could be considered an
increase in the ``effective optical depth'' of the corona. The BLR is
then deprived of the ionizing photons, so the emission-line and UV
continuum variability are decoupled (the ``anomaly''). The increase in
the effective optical depth in the anomaly spectrum is $\Delta \tau =
1.8$, {\bf implying a reduction in the EUV flux by approximately
  $e^{-\Delta \tau}$ or $16.5\%$. Interestingly, the observed deficit in
  the \civ broad emission line flux during the anomaly is of a similar
  amplitude (Fig 2 and Fig. 1e in Paper IV)}.

Thus, the warm Comptonization model can simultaneously explain
all three observations (the UV emission-line flux decrease, the
soft-excess increase, and the emission-line anomaly), so it is our
preferred model. We do not understand why the corona may change its
physical structure in this way; observations such as these provide
motivations for further theoretical work on the structure of the
accretion disk and the corona.  Notably, an anomalous continuum behavior
has been seen before. In 3C~273, there was one epoch of \xmm observations
when the UV flux decreased but the X-ray flux increased (Page et
al. 2004), while the source otherwise behaved normally.




\section{Conclusion}

In this paper, we report on analyses of \swift and \chandra X-ray spectra
taken during the \HST monitoring campaign of NGC 5548 that help us
understand the UV anomaly reported in Paper IV. We show that obscuration
of the continuum source is unlikely to be the cause of the
anomaly. Instead, the spectral energy distribution of the continuum
changed during the anomaly, as seen in the X-ray spectra. A possible
scenario may be that the warm Comptonizing corona covered more of the
accretion disk during the anomaly, depleting EUV photons while
increasing the soft X-ray excess. The decrease in the ionizing continuum then
leads to the emission-line anomaly. In order to understand the finer
details of the anomaly, detailed photoionization models will be
necessary. These results demonstrate the importance of contemporaneous
X-ray spectra to interpreting high-quality RM data.  We suggest that
future RM campaigns in the optical and/or UV include
an X-ray component as well. \\

\noindent {\bf Acknowledgments:} We are grateful to Christine Done for
discussions on the warm Comptonization model. Support for this work was
provided by the National Aeronautics and Space Administration (NASA)
through Chandra Award Number G04-15114X to S.M. issued by the Chandra
X-ray Observatory Center, which is operated by the Smithsonian
Astrophysical Observatory for and on behalf of NASA under contract
NAS8-03060. Support for {\it HST} program GO-13330 was provided by NASA
through a grant from the Space Telescope Science Institute, which is
operated by the Association of Universities for Research in Astronomy,
Inc., under NASA contract NAS~5-26555. C.S.K. is supported in part by
NSF grants AST-1515876 and AST-1515927.  K.H. acknowledges support from
STFC grant ST/M001296/1.  K.L.P. and P.A.E. ackowledge support from the
UK Space Agency.  A.V.F.'s group at UC Berkeley is grateful for
financial assistance from NSF grant AST-1211916, the TABASGO Foundation,
and the Christopher R.  Redlich Fund. Y.K. acknowledges support from the
grant PAIIPIT IN104215 and CONACYT grant168519.  This work made use of
data supplied by the UK Swift Science Data Centre at the University of
Leicester (see Evans et al.  2009).

\clearpage

\begin{table}
\scriptsize
\caption{\swift exposure times}
\begin{tabular}{ccc}
\hline
Days & Julian dates  & Exposure Time (ks)\\
\hline
 0--55  & 2,456,690.4189 -- 2,456,744.5088   & 58 \\
 55--75 &  2,456,745.3676 -- 2,456,764.7642   & 32 \\
 75--85 &  2,456,765.6905 -- 2,456,775.5760   & 7  \\
 85--100 &   2,456,776.0284 -- 2,456,790.6312  & 20  \\
 100--110 & 2,456,791.0273 -- 2,456,800.3587      &  12 \\
  110--120 & 2,456,800.8177 -- 2,456,810.5546     &  14  \\
  120--135 &  2,456,810.8812 -- 2,456,825.5525    & 10  \\
  135--150 & 2,456,825.9454 -- 2,456,840.2837       &  16  \\
  150--170 & 2,456,840.2768 -- 2,456,859.7580 &  15  \\
\hline-
\end{tabular}
\end{table}

\begin{table}
\scriptsize
\caption{Fits to \swift spectra: absorbed power-law plus black-body model$^{1}$}
\begin{tabular}{lccccccc}
\hline
Obs ID  &  BB~($kT$)$^2$ & BB~Norm & BB~Flux & Intrinsic Absorption & Photon Index$^2$ & Power-law Norm$^2$\\
        &   keV   & $10^{-5}$ & $0.1$--$2$ keV & $N_{\rm H}$   & $\Gamma$ &    $10^{-3}$    \\
 && ph~keV$^{-1}$~s$^{-1}$~cm$^{-2}$ & $10^{-12}$ ergs cm$^{-2}$ s$^{-1}$ & $10^{22}$~cm$^{-2}$ & & ph~keV$^{-1}$~s$^{-1}$~cm$^{-2}$\\
\hline
Anomaly   &    $0.12\pm0.005$    & $4.5\pm0.2$ & 3.7  & $0.66\pm0.05$    &  $1.49\pm0.04$   &  $5.9\pm0.3$  \\
Pre-anomaly   &    $0.12\pm0.005$    & $1.7\pm0.1$ & 1.4  & $1.13\pm0.06$    &  $1.49\pm0.04$   &  $5.9\pm0.3$  \\
\hline
\end{tabular}
1. A scattered component is also included in the fit (see text). \\
2. Power-law parameters and BB temperature are tied for both the datasets. \\
3. $\chi^2_{\nu}$ for the joint fit is 1.16 for $\nu=1001$ degrees of freedom. 
\end{table}

\begin{table}
\scriptsize
\caption{\swift spectra: warm Comptonization fit parameters$^{1,2}$}
\begin{tabular}{lccc}
\hline
Spectrum  &  Optical depth   & Normalization \\
                  &    &     \\
\hline
Anomaly   &   $22.34\pm0.16$   &  $50.34\pm 1.85$        \\
Pre-anomaly  &   $20.55\pm0.15$ &  $22.09\pm 0.94$        \\
\hline
\end{tabular}

\noindent
1. Power-law parameters are tied for both the datasets.  \\
2. The warm corona temperature was fixed at 0.15 keV and the seed photon 
temperature to 0.74 eV. Only one parameter (optical depth or normalization) 
was allowed to vary at a time. \\
\end{table}

\begin{table}
\scriptsize
\caption{\swift spectra: PHASE fit parameters}
\begin{tabular}{lcccc}
\hline
Spectrum  &  Log $U$  &  Log ($N_{\rm H}$/cm$^{-2}$)   & Photon Index$^1$   \\
        &          &    &  $\Gamma$    \\
\hline
Anomaly   & $-0.3\pm0.03$ &  $22.12\pm0.01$  &   $1.42\pm0.02$        \\
Pre-anomaly  & $-0.4\pm0.01$ &  $22.18\pm0.01$  &   $1.42\pm0.02$         \\
\hline
\end{tabular}

\noindent
1. Power-law parameters are tied for both the datasets. \\
2. $\chi^{2}_{\nu}$ for the joint fit is 1.1 for $\nu=1003$ degrees of freedom. 
\end{table}

\begin{table}
\scriptsize
\caption{NGC~5548 Chandra Observation Log}
\begin{tabular}{lcccc}
\hline
Obs.&  ID & Date of Observation & JD & Exposure Time \\
\hline
I & 15659  & 2014 Feb. 3   & 2,456,692  (pre-anomaly)   &   5 ks \\
II & 15660  & 2014 Mar 30  & 2,456,747  (just before anomaly)  &   5 ks \\
III & 15661  & 2014 May 6   & 2,456,784  (during anomaly)   &   5 ks \\
IV & 15662  & 2014 June 23  & 2,456.832  (post-anomaly)  &   5 ks \\
\hline
\end{tabular}
\end{table}

\clearpage

\begin{figure}
\begin{center}
\includegraphics[width=10cm]{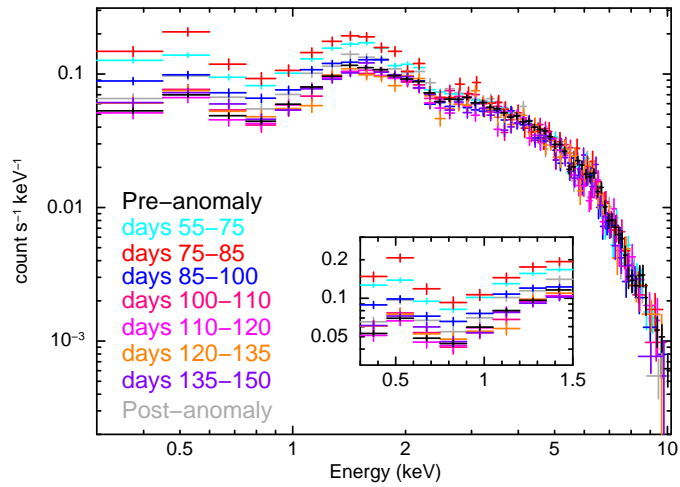}
\end{center}
\caption{\swift spectra in different time bins. Note the increase in the
  soft excess from the pre-anomaly phase, peaking at days 75--85 (red)
  and then decreasing. This is a model-independent result, and suggests
  that the change in the spectral energy distribution of NGC~5548 is
  responsible for the anomaly. The inset shows $0.3$ to $1.5$ keV spectra 
  for clarity.}
\end{figure}

\begin{figure}
\begin{center}
\includegraphics[width=10cm]{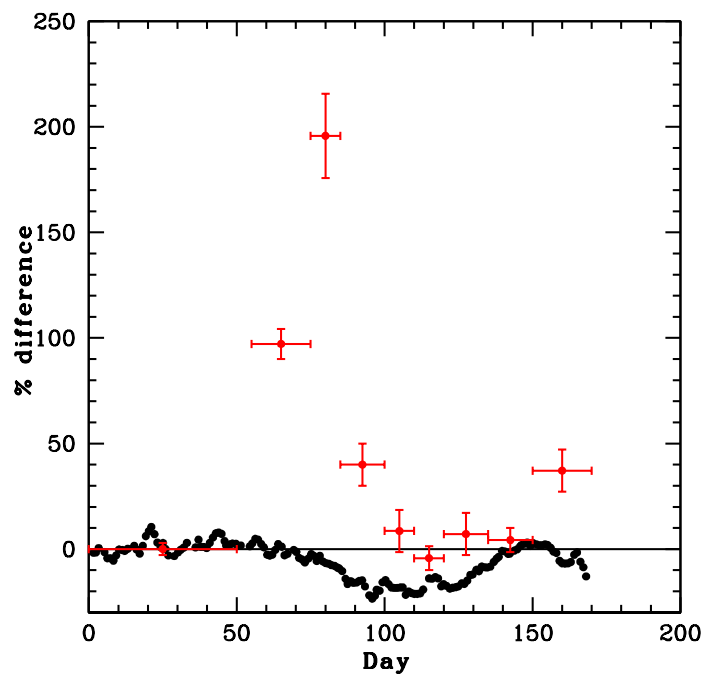}
\end{center}
\caption{NGC~5548 light-curve. The black points show the percentage
  deficit in the \civ flux (reproduced from Fig. 1e of Paper IV). This
  shows the onset of the anomaly around day 75 of the campaign. The red
  points show the percentage excess in the \swift count rate at $\approx
  0.55$ keV (as in Fig. 1). We see that the soft excess increases before
  the start of the anomaly, peaks during the period of anomaly and then
  declines. }
\end{figure}

\begin{figure}
\begin{center}
\includegraphics[width=7cm]{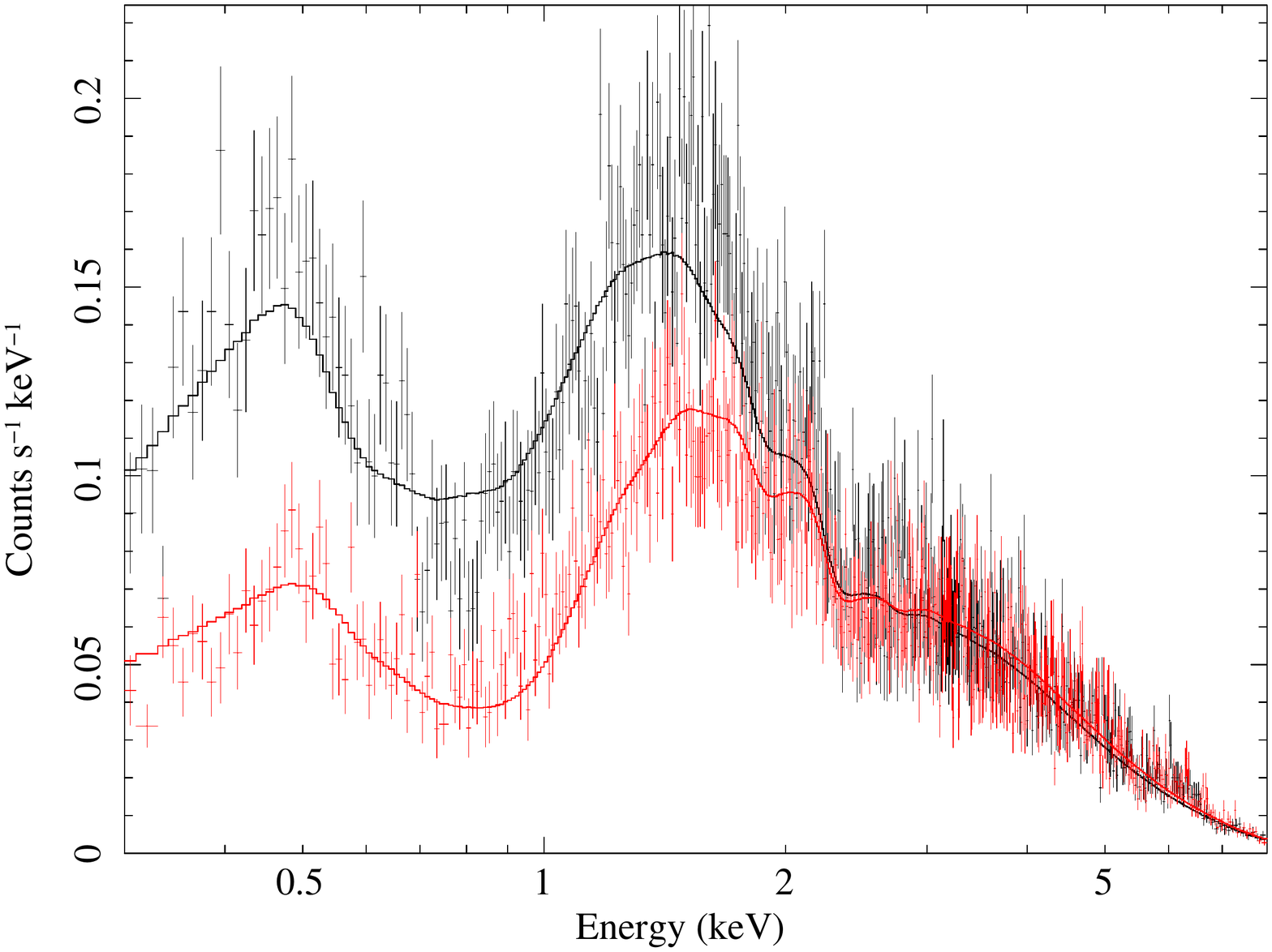}
\includegraphics[width=7cm]{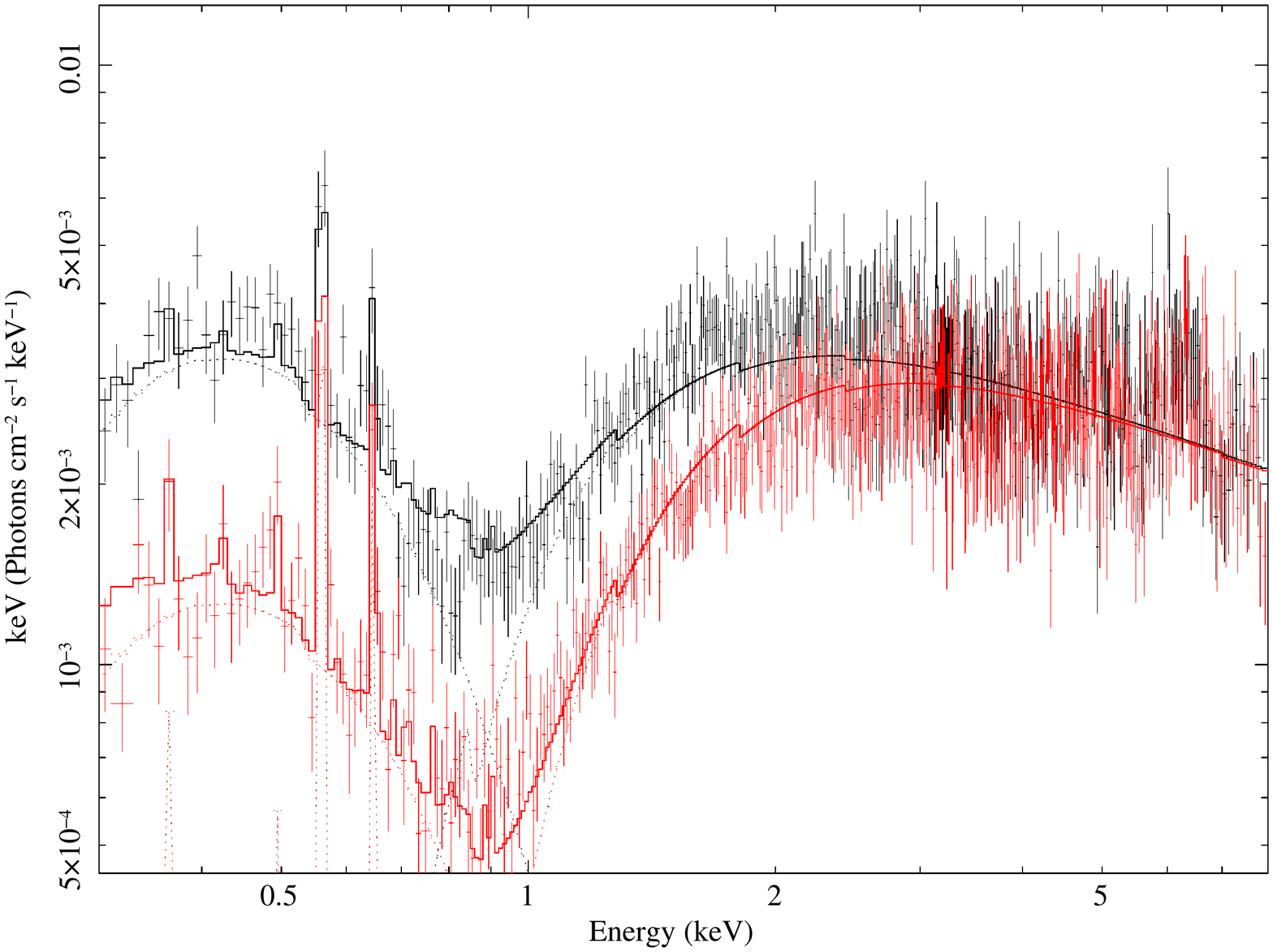}
\end{center}
\caption{{\it Top:} \swift spectra: anomaly (black) and pre-anomaly
  (red) fit with an absorbed power law plus a black-body model. An
  additional scattered component is also included.  {\it Bottom:} The
  {\bf $Ef(E)$} intrinsic spectra (without folding in the instrument
  response); the three model components are shown as dotted lines. }
\end{figure}

\begin{figure}
\begin{center}
\includegraphics[width=7cm]{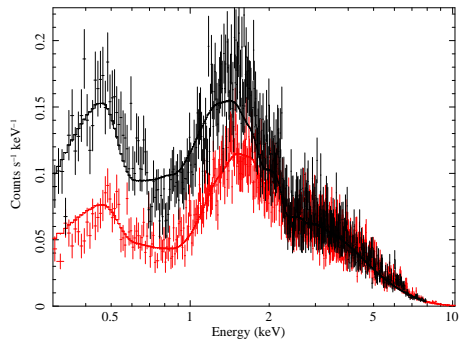}
\end{center}
\caption{As in Fig. 3, but fit with a warm Comptonization model.  
}
\end{figure}

\begin{figure}
\begin{center}
\includegraphics[width=7cm]{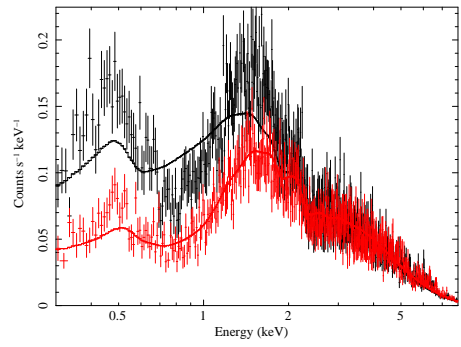}
\end{center}
\caption{The \swift spectra fitted with a partial covering absorber
  model. This is clearly a poor fit compared to the models shown in
  Figs. 3 and 4.  }
\end{figure}

\clearpage

\begin{figure}
\begin{center}
\includegraphics[width=10cm,height=10cm]{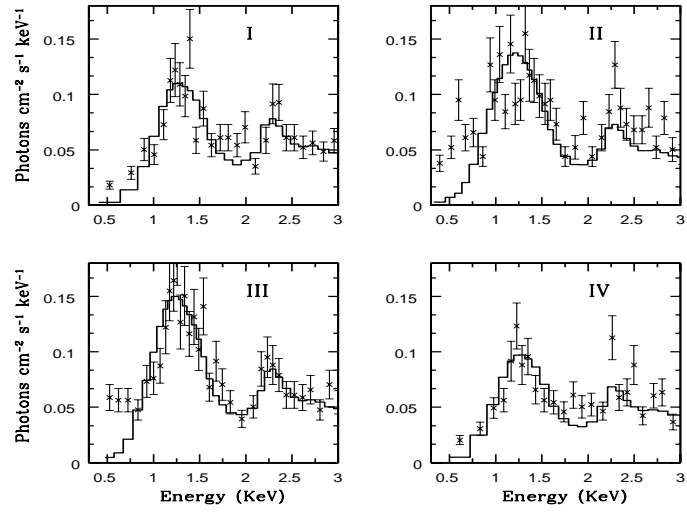}
\end{center}
\caption{\chandra spectra with the best-fit absorbed power-law model. The 
  soft excess is apparent in observations II and III.  }
\label{fig5}
\end{figure}

\end{document}